\documentclass[conference]{IEEEtran}
\IEEEoverridecommandlockouts
\usepackage{cite}
\usepackage{amsmath,amssymb,amsfonts}
\usepackage{algorithmic}
\usepackage{graphicx}
\usepackage{textcomp}
\usepackage{xcolor}
\usepackage[font=small]{caption}
\usepackage{subcaption}

\usepackage{acronym}

\newacro{gfm}[GFM]{Grid-Forming}
\newacro{gfl}[GFL]{Grid-Following}
\newacro{ibr}[IBR]{Inverter-Based Resource}
\newacro{pll}[PLL]{Phase-Locked Loop}
\newacro{vsm}[VSM]{Virtual Synchronous Machine}
\newacro{voc}[VOC]{Virtual Oscillator Control}
\newacro{sm}[SM]{Synchronous Machine}
\newacro{cf}[CF]{Complex Frequency}
\newacro{coi}[CoI]{Center of Inertia}
\newacro{dfig}[DFIG]{Doubly-Fed Induction Generator}
\newacro{csig}[CSIG]{Constant Speed Induction Generator}
\newacro{smib}[SMIB]{Single-Machine-Infinite-Bus}
\newacro{bess}[BESS]{Battery Energy Storage Systems}
\newacro{facts}[FACTS]{Flexible Alternating Current Transmission System}
\newacro{wac}[WAC]{Wide-Area Control}

\def\BibTeX{{\rm B\kern-.05em{\sc i\kern-.025em b}\kern-.08em
    T\kern-.1667em\lower.7ex\hbox{E}\kern-.125emX}}
\begin{document}

\title{Coherency Control in Power Systems
\thanks{%
  \noindent
  This work is supported by Sustainable Energy Authority of Ireland (SEAI) by funding R.~Bernal, I.~Ponce and F.~Milano under project FRESLIPS, Grant No.  RDD/00681.%
}
}

\author{
\IEEEauthorblockN{Rodrigo Bernal, Ignacio Ponce and Federico Milano}
\IEEEauthorblockA{School of Electrical \& Electronic Engineering \\
University College Dublin,
Dublin, Ireland\\
rodrigo.bernal@ucdconnect.ie,
ignacio.poncearancibia@ucdconnect.ie,
federico.milano@ucd.ie\vspace{-4mm}}
}

\maketitle

\begin{abstract}
This paper proposes a coherency control strategy for \acp{ibr} to establish coherence among power system devices.  Using the equivalence of the \ac{cf} of the injected currents as the definition for coherency among devices, the control enforces an output current with a proportional magnitude and a constant phase shift relative to a reference.  This formulation makes the control technology-agnostic, enabling coherency with any type of resource.  Case studies based on the two-area and IEEE 39-bus systems demonstrate the controller's potential to improve damping and overall dynamic behavior.  The paper further evaluates practical implementation aspects including delay/noise sensitivity and the trade-off between oscillation mitigation and disturbance propagation.  This work establishes coherency as a viable direct control objective for \acp{ibr} in modern power systems.
\end{abstract}


\vspace{-1mm}

\section{Introduction}
\subsection{Motivation}
Coherency traditionally refers to the tendency of \acp{sm} to maintain their relative rotor angles during a transient—a condition often described as ``oscillating together'' \cite{chow}.  This coherent behavior is potentially beneficial to system stability, as it reduces the risk of angular separation, mitigates inter-area oscillations, and supports the operation of coordinated control schemes \cite{kundur}.  These properties suggest that explicitly designing control architectures to achieve coherency could be advantageous, particularly given the flexibility of \acp{ibr}.  However, this idea has remained relatively unexplored in modern grids due to the lack of a suitable definition of coherency applicable beyond \acp{sm}.  
Leveraging on recent work by the author \cite{coherencydef}, this paper presents a coherency control strategy for \acp{ibr} capable of regulating its degree of coherency with respect to a device chosen as reference, serving as a proof of concept for more advanced coherency-based control designs.


\subsection{Literature review}

Research on \ac{wac} schemes has exploited coherency to design inter-area oscillation damping controllers.  A relevant thread is sparsity-promoting optimal control, which simultaneously identifies sparsity patterns and introduces feedback to penalize relative motion between coherent areas \cite{sparsity1, sparsity2, sparsity3}.  A foundational work has shown that coherency can aid the selection of measurement sites to complement and improve the performance of power system stabilizers \cite{hquebec}.  Subsequent research generalized this approach to a coordinated \ac{wac} on \ac{facts} devices \cite{wacfacts}.  A complimentary research line proposes a systematic procedure to guide the selection of input/output signals for \acp{wac} based on a pre-identification of coherent groups of \acp{sm} \cite{padhy}.  In summary, existing coherency-based \ac{wac} architectures assume that coherent areas already exist in the system, either predefined or identified online.  However, how generators are grouped in transient conditions depends on the disturbance \cite{chow}. 
As controllers are designed and tuned assuming certain coherent clusters, this dependency can impact negatively their performance.

This challenge has motivated a second line of research aimed at actively enforcing coherency among \acp{sm}.  For instance, consensus approaches for cyber-physical power systems have demonstrated that coordinated feedback under communication constraints can restore synchronism and enhance rotor-angle stability \cite{yao2019consensus}.  Second-order consensus theory has been used to design control protocols where a group of \acp{sm} designated as followers achieve coherent motion with a single leader machine \cite{yingjun2014}.  Another study has proposed a coherency-based control that explicitly drives machines to oscillate along a reference trajectory to improve dynamic security \cite{DETUGLIE20081425}.  However, these existing coherency-based control strategies remain limited to \acp{sm}.

A more general definition of coherency, recently established using the \ac{cf} concept \cite{coherencydef}, provides a broader formulation.  This paves the way to exploiting coherency as a control objective for \acp{ibr}, an ideal feature for modern grids composed of heterogeneous mixes of generation technologies.



\subsection{Contribution}
The primary contribution of this paper is a coherency controller for \acp{ibr} capable of regulating its coherence with a remote reference.  The control strategy is formulated using only measurable terminal quantities, making it independent from specific device models.  A secondary contribution is an analysis of how this enforced coherency influences overall power system dynamics.

\subsection{Paper organization}
The remainder of this document is organized as follows.  Section \ref{sec:framework} provides the background on \ac{cf}-based coherency framework. 
Section \ref{sec:framework} details the implementation of a coherency control for \acp{ibr}.
Section \ref{sec:casestudies} presents case studies to illustrate the potential of this approach for improving the system's dynamic response.  Finally, Section \ref{sec:conclusion} concludes the paper and outlines future research lines.

\section{Coherency Framework}\label{sec:framework}
In this section, relevant background on modern coherency theory on which the proposal is founded is presented.
\subsection{Complex frequency}
Consider a time-varying complex quantity in polar coordinates $\overline{x}(t)=x(t)\,e^{\jmath\,\phi(t)}$.  The \ac{cf} of $\overline{x}$ is denoted as $\overline{\eta}_{x}(t)$ and defined as follows \cite{cf}:
\begin{equation}
    \overline{\eta}_{x}(t)=\frac{\dot{x}(t)}{x(t)}+\jmath\,\dot{\phi}(t)\,= \rho_x+\jmath\,\omega_x,
\end{equation}
where $\cdot$ denotes the time derivative.

The imaginary part of the \ac{cf} ($\omega_x$) is the time derivative of the phase, representing the conventional instantaneous angular frequency.  The real part ($\rho_x$), defined by magnitude variations, represents the instantaneous bandwidth.



\subsection{Ideal instantaneous coherency}
A novel definition of instantaneous coherency in terms of the \ac{cf} has been introduced recently in \cite{coherencydef}.  According to this definition, two complex quantities $\overline{x}(t)$, $\overline{y}(t)$ are perfectly coherent if their \acp{cf} are equal:
\begin{equation}
    \overline{\eta}_{x}(t)=\overline{\eta}_y(t)\,.
\end{equation}

In the electromechanical time scale, ideal coherency among devices is reflected on an equality in the \ac{cf} of their current injections to the network \cite{coherencydef}.  Formally, considering two devices denoted by subscripts $1$ and $2$ and their current injections $\overline{\imath}_{1}$, $\overline{\imath}_2$, they achieve ideal coherency if:
\begin{equation}\label{eq:cc}
    \overline{\eta}_{\imath_1}(t)=\overline{\eta}_{\imath_2}(t) \, .
\end{equation}



The ideal coherency condition (\ref{eq:cc}) has been introduced in terms of time derivatives, which can be difficult to implement in terms of control.  Nevertheless, this condition also implies that the magnitude of the currents is proportional and their phase has a constant difference.  This is derived from integrating (\ref{eq:cc}), as follows:
\begin{align}
\int \overline{\eta}_{\imath_1}dt &=  \int\overline{\eta}_{\imath_2}dt, \notag\\
\int \left( \rho_{\imath_1} + \jmath\omega_{\imath_1} \right) dt &= \int \left( \rho_{\imath_2} + \jmath\omega_{\imath_2} \right) dt, \notag\\
\int \left( \frac{\dot{\imath}_1}{\imath_1} + \jmath \dot{\theta}_{\imath_1} \right) dt &=  \int \left( \frac{\dot{\imath}_2}{\imath_2} + \jmath \dot{\theta}_{\imath_2} \right) dt, \notag\\
\ln{\imath_1} + \jmath \theta_{\imath_1}   &=  \ln{\imath_2} + \jmath \theta_{\imath_2} + \overline{k}, \label{eq_int_curr}
\end{align}
where $\overline{k}$ is a complex constant that arises from the integral initial conditions and where we have omitted the dependency on time for simplicity.  By taking the exponential on each side of (\ref{eq_int_curr}) and separating the expression in their real and imaginary part, one has:
\begin{align}
\imath_1 &=k_{\imath}\,\imath_2,\label{eq_curr_prop_real} \\
\theta_{\imath_1} &=\theta_{k_{\imath}}+\theta_{\imath_2}, 
\label{eq_curr_prop_imag}
\end{align}
where $k_{\imath}$ is a proportional constant that relates the magnitudes of the currents and $\theta_{k_{\imath}}$ is the constant phase difference between current phases, both derived from the complex constant $\overline{k}$.

\begin{figure}
    \centering
    \includegraphics[width=0.9\linewidth]{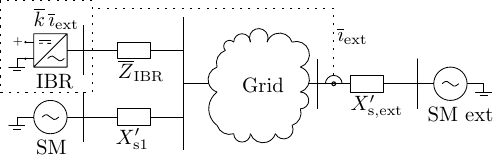}
    \caption{Simplified diagram of coherency control.}
    \label{fig:diagram}
\end{figure}

The coherency control is implemented for a power-controlled \ac{ibr}.  The controller generates power references by multiplying the terminal voltage with a current derived from an external measurement, $\overline{\imath}_{\mathrm{ext}}$, scaled by a complex gain $\overline{k}$ (equations \eqref{eq_curr_prop_real}-\eqref{eq_curr_prop_imag}).  These references are then fed into the standard grid-following (GFL), power-based control of the \ac{ibr}.  Figure~\ref{fig:diagram} illustrates, with a simple diagram, the proof of concept of the proposed coherency control.

To enable a smooth transition between coherent and independent dynamics, we introduce a coherency parameter, $C$.  This parameter defines the proportion of a device's power output that is forced to be coherent with a predetermined generator.  For example, when $C=0.5$ is applied to a synchronous machine, 50\% of its power is delivered using its inherent machine dynamics (effectively representing half of its original nominal capacity), while the remaining 50\% is delivered by a controlled current source—implemented in practice by an \ac{ibr}—whose dynamics are governed with \eqref{eq_curr_prop_real}-\eqref{eq_curr_prop_imag}.  In this hybrid structure, local voltage control and synchronization are maintained with the synchronous machine portion, while the coherency control acts as a distributed power source that follows an external reference.

\section{Case Studies}\label{sec:casestudies}
In this section, two case studies are presented: the two-area Kundur system and the 39-bus system.  The first system is used to study the overall impact of coherency control on inter-area oscillations, including the effects of increasing the share of coherency control, the implications of communication delays, noise, and the results of using only the imaginary part of the coherency function for control.  The second system is used to evaluate the impact of enforcing a predetermined coherent cluster of devices through coherency control.
All \ac{ibr} that are added are modeled with coherency control.  This model implements a \ac{gfl} with PLLs serving for synchronization with parameters $K^{\mathrm{PLL}}_p=0.1$ and $K^{\mathrm{PLL}}_i=0.05$, algebraic voltage loops and current loops that are simplified as first order delays with time constant $\tau_{\imath_{\mathrm{dq}}}=0.01$s.
\subsection{Two-area Kundur System}
The Kundur two-area system is a standard test case for analyzing both local and inter-area power oscillations.  The system comprises four \acp{sm} separated into two areas, interconnected by a high-impedance transmission line.  Two machines (\acp{sm} 1 and 2) are in one area, and the other two (\acp{sm} 3 and 4) are in the second area.

The \ac{sm} models are 6th-order, equipped with simplified IEEE DC-1 automatic voltage regulators (AVRs) and Type I turbine governors.  Line dynamics is neglected and loads are considered as constant impedance. 

\subsubsection{Dynamic Effect of Increasing Coherency Among Devices} \label{sec:increasing_C}
This subsection presents a brief analysis of the dynamic effects of increasing the share of coherency control among devices.  A \ac{gfl} \ac{ibr} is placed at buses 1, 2, and 4, operating in parallel with the \acp{sm} at those locations.  A coherency control, designed to imitate the current dynamics of \ac{sm} 3, is added to each \ac{ibr}.

\begin{figure}[htbp]
    \centering
    \begin{minipage}{0.5\linewidth}
        \centering
        \includegraphics[width=\linewidth]{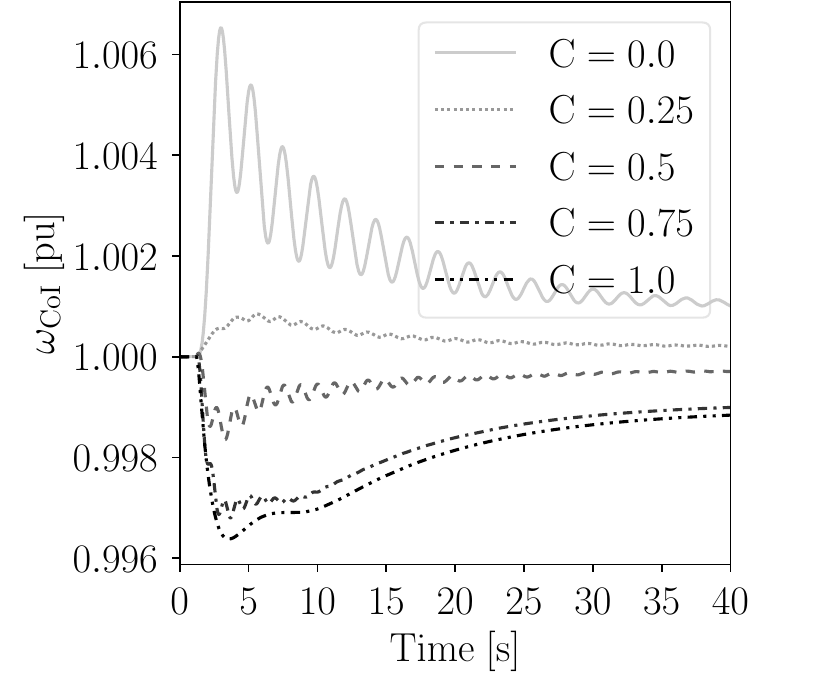}
        \subcaption*{(a)} 
    \end{minipage}\hfill
    \begin{minipage}{0.5\linewidth}
        \centering
        \includegraphics[width=\linewidth]{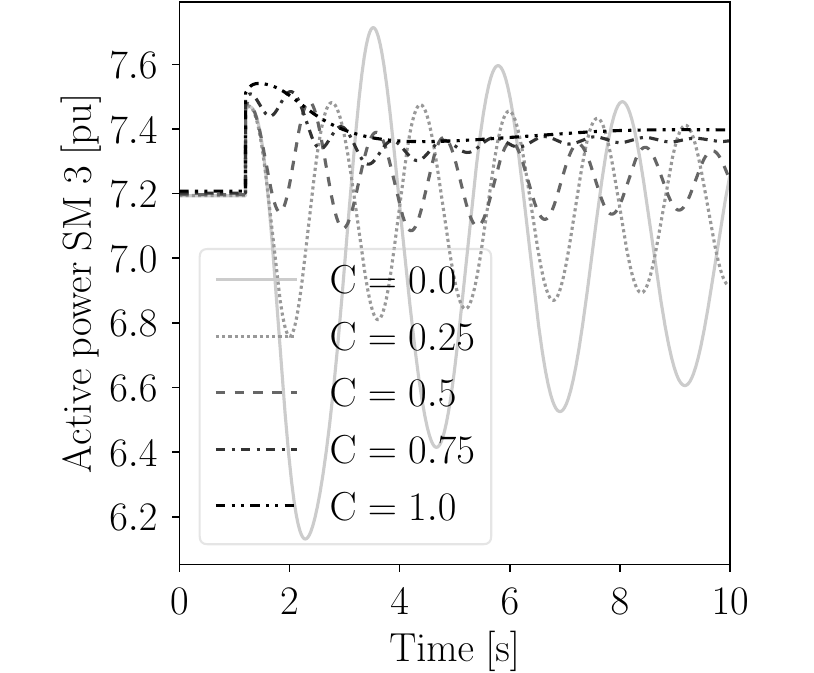}
        \subcaption*{(b)} 
    \end{minipage}
    \caption{Two-area system – Loss of one circuit of the line connecting buses 7 and 8. Frequency of the \ac{coi} (a), and active power (b).}
    \label{fig:Cperc}
\end{figure}

Figure~\ref{fig:Cperc} illustrates the impact on the frequency of the \ac{coi} (panel a) and the active power injected by \ac{sm} 3 (panel b).  The results are shown as the coherency control share, $C$, is increased from 0 to 1 across all \acp{ibr} for the loss of one circuit on the line connecting buses 7 and 8 (increasing the impedance between the two areas).  The parameter $C$ represents the share of the control that emulates the dynamics of \acp{sm} 1, 2, and 4, as defined in Section~\ref{sec:framework}.

A clear impact is observed on the active power oscillations between \ac{sm} 3 and the rest of the system.  As the parameter $C$ increases, the frequency of the oscillations increases while their amplitude is reduced.  Consequently, oscillations in the frequency of the \ac{coi} are mitigated when the coherency control share is maximized.  Furthermore, in this case, the entire system behaves as a single, larger generator, with its power distributed through the \acp{ibr}.  

\subsubsection{Sensitivity to Delays and Noise} \label{sec:delays_noise}
In this subsection, we assess non-ideal conditions for the coherency control by studying the effects of communication delays and noise in the measured currents, which are used as input signals.

\begin{figure}[htbp]
    \centering
    \begin{minipage}{0.5\linewidth}
        \centering
        \includegraphics[width=\linewidth]{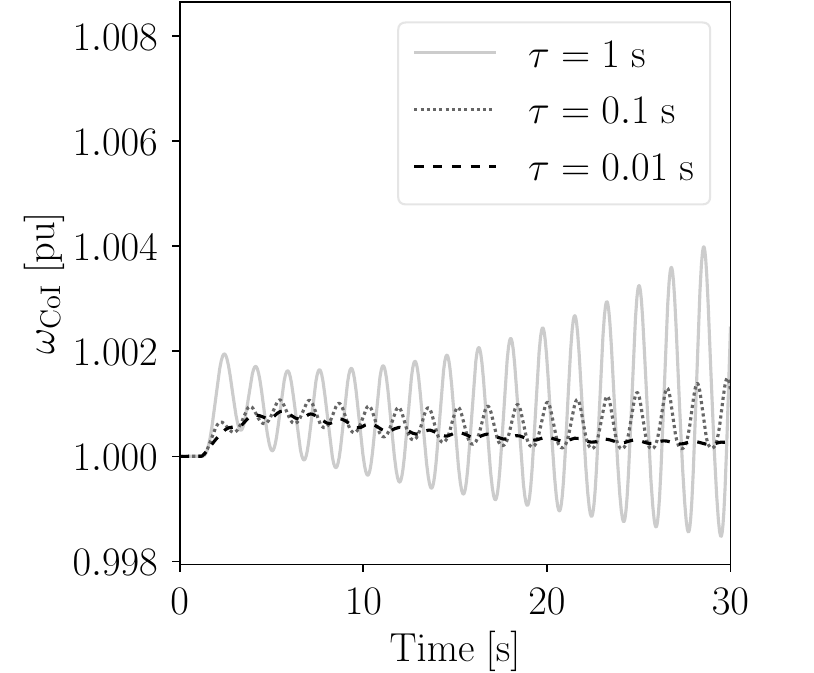}
        \subcaption*{(a)} 
    \end{minipage}\hfill
    \begin{minipage}{0.5\linewidth}
        \centering
        \includegraphics[width=\linewidth]{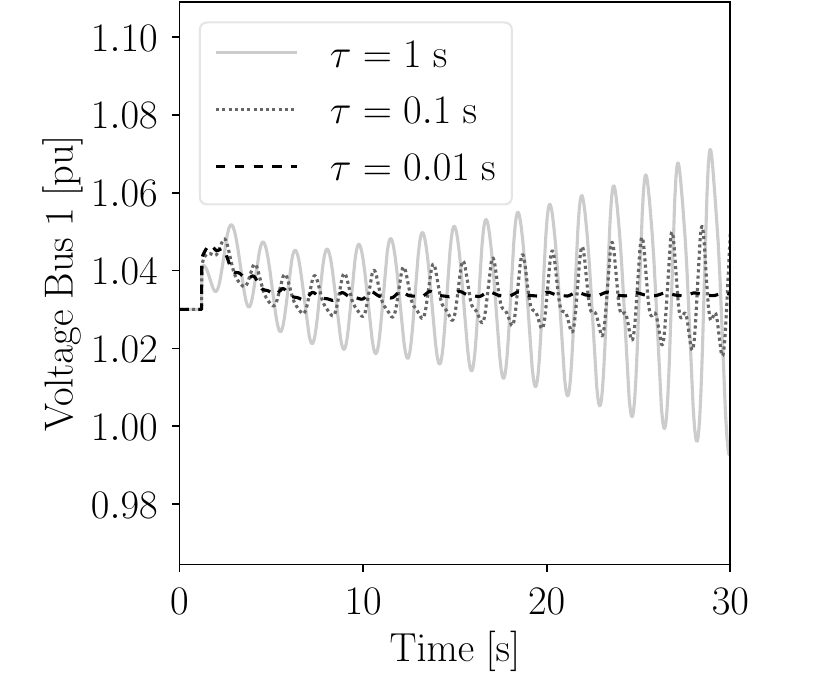}
        \subcaption*{(b)} 
    \end{minipage}
    \caption{Two-area system – Loss of one circuit of the line connecting buses 7 and 8 - $\mathrm{C}=0.25$.  Frequency of the \ac{coi} (a), and voltage at bus 1 (b) for different delays in the current input signal.}
    \label{fig:Delay}
    \vspace{-3mm}
\end{figure}

Figure~\ref{fig:Delay} shows the frequency of the \ac{coi} (panel a) and the voltage at bus 1 (panel b) for the loss of one circuit of the line connecting buses 7 and 8.  The results are shown for communication delays of 0.01 s, 0.1 s, and 1 s, modeled using a first-order approximation in the measured current signal.  Using the same topology as in Section~\ref{sec:increasing_C}, we study the case for which $\mathrm{C}=0.25$ for all \acp{ibr}.

\label{sec_delay}

Longer time delays are more likely to induce system instability, particularly when the dominant oscillation modes occur on a time scale comparable to the delay itself.  This represents a physical constraint of the coherency control strategy and makes its implementation feasible to clusters of devices where communication distances—and thus delays—remain within suitable operation conditions.  
This results is also in line with the intuition that devices that are geographical close are more likely to belong to the same coherency cluster.

\begin{figure}[htbp]
    \centering
    \begin{minipage}{0.5\linewidth}
        \centering
        \includegraphics[width=\linewidth]{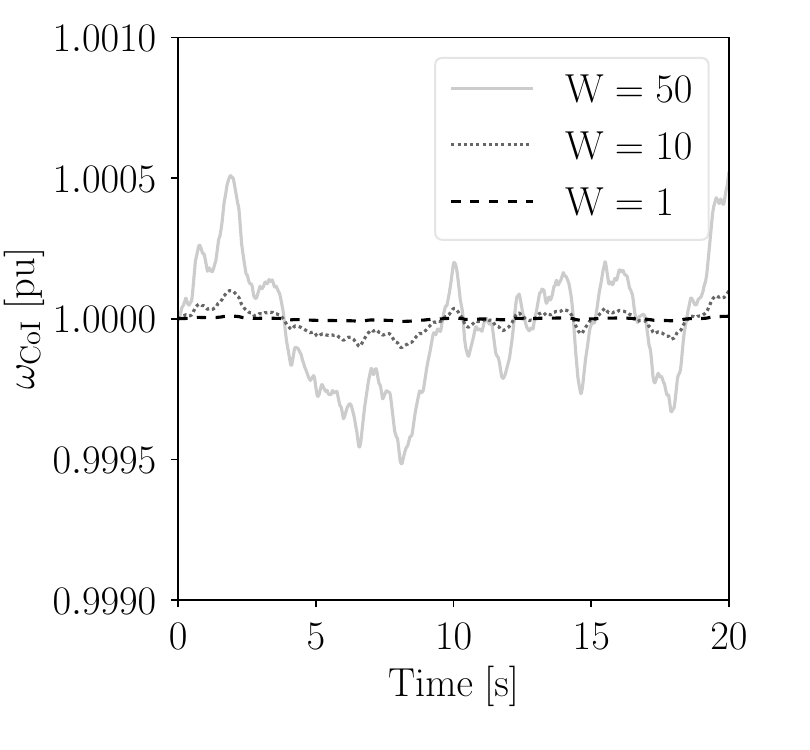}
        \subcaption*{(a)} 
    \end{minipage}\hfill
    \begin{minipage}{0.5\linewidth}
        \centering
        \includegraphics[width=\linewidth]{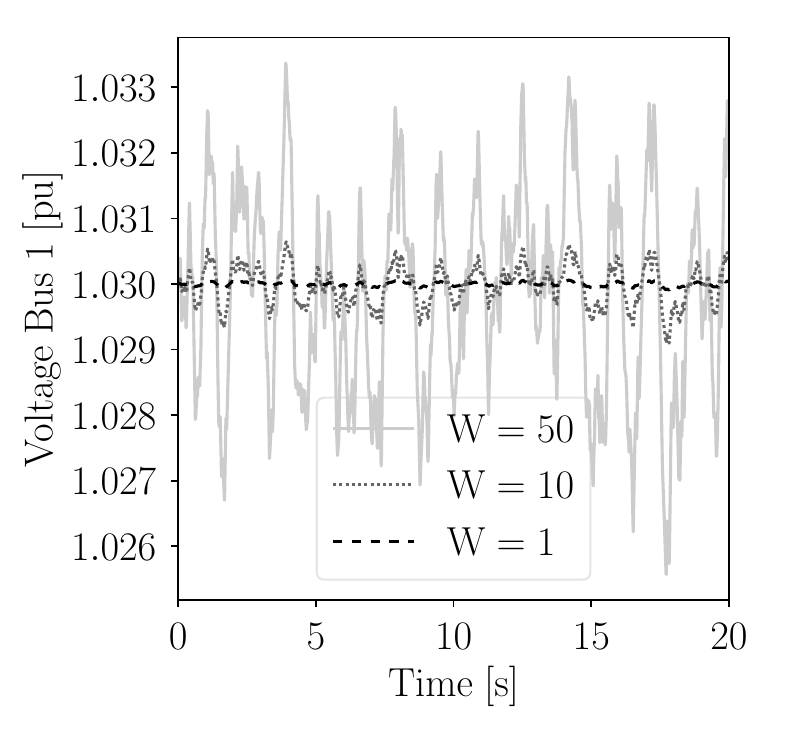}
        \subcaption*{(b)} 
    \end{minipage}\vfill
    \begin{minipage}{0.5\linewidth}
        \centering
        \includegraphics[width=\linewidth]{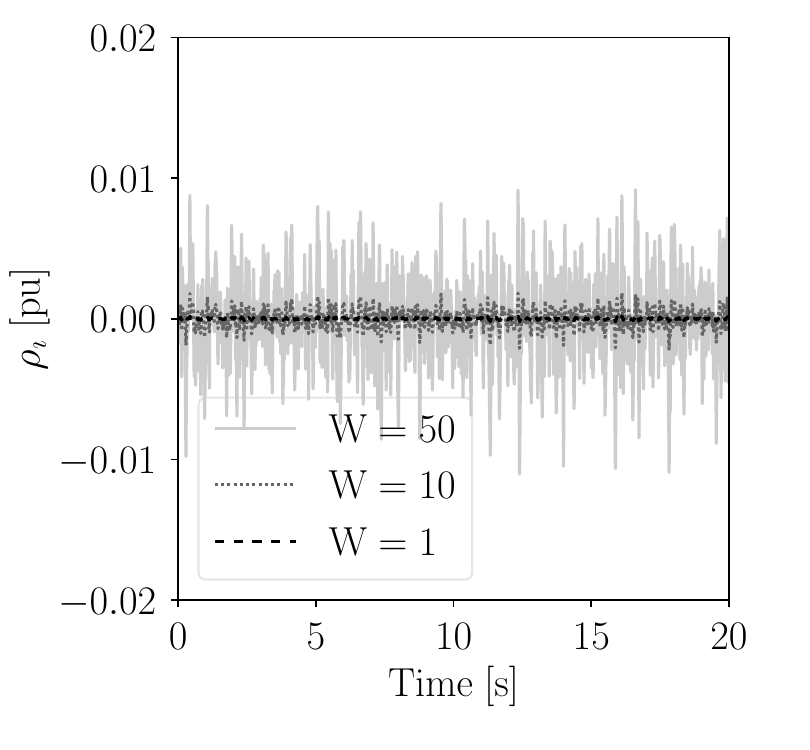}
        \subcaption*{(c)} 
    \end{minipage}\hfill
    \begin{minipage}{0.5\linewidth}
        \centering
        \includegraphics[width=\linewidth]{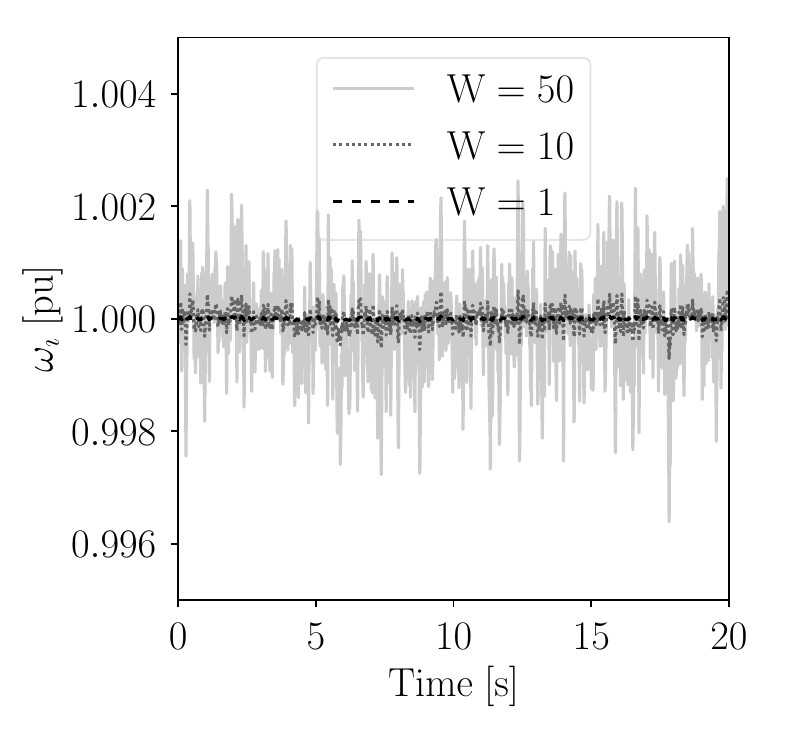}
        \subcaption*{(d)} 
    \end{minipage}
    \caption{Two-area system – Noise in the measured signal with weighting factor $W = 1$, 10 and 50 - $\mathrm{C}=0.25$ - Frequency of the \ac{coi} (a), voltage at bus 1 (b), real (c) and imaginary (d) parts of the measured \ac{cf} of the current injected at bus 3.}
    \label{fig:noise}
    \vspace{-2mm}
\end{figure}

To evaluate the impact of measurement noise, a normally-distributed (Ornstein-Uhlenbeck) process with a standard deviation of $\sigma=1\%$ and an autocorrelation parameter $\alpha=10$ is directly added to the measured current input.  This noise is then scaled by a weighting factor, $W$.  Figure~\ref{fig:noise} illustrates the effects of varying this weighting factor at values of 1, 10, and 50. The W=50 case represents an overly extreme scenario, serving as a conservative example to demonstrate noise propagation through the system.  Panels (a) and (b) present the frequency of the \ac{coi} and the voltage at bus 1, respectively, while panels (c) and (d) show the noise's effect on the real and imaginary parts, respectively, of the measured output \ac{cf} of the injected current at bus 1.

As the noise level increases, its negative effect---propagated by the coherency control throughout the entire system---becomes evident.  Even without contingencies, the noise naturally excites oscillation modes, whose frequency is clearly observable in the voltage at bus 1.  In the worst case ($W=50$), this leads to voltage deviations of up to 0.05 pu at this node. 

In cases where noise is significant, proper filtering is required to avoid propagation of unwanted oscillatory behavior.  Nevertheless, a careful trade-off must be made between the level of filtering and the delay introduced by the filter.  This is essential to avoid inducing stability issues, such as the one observed in Figure~\ref{fig:Delay}.

\subsubsection{Conventional vs Complex Coherency} 
\label{sec:conv_vs_cmplx}

\begin{figure}[htbp]
    \centering
    \begin{minipage}{0.5\linewidth}
        \centering
        \includegraphics[width=\linewidth]{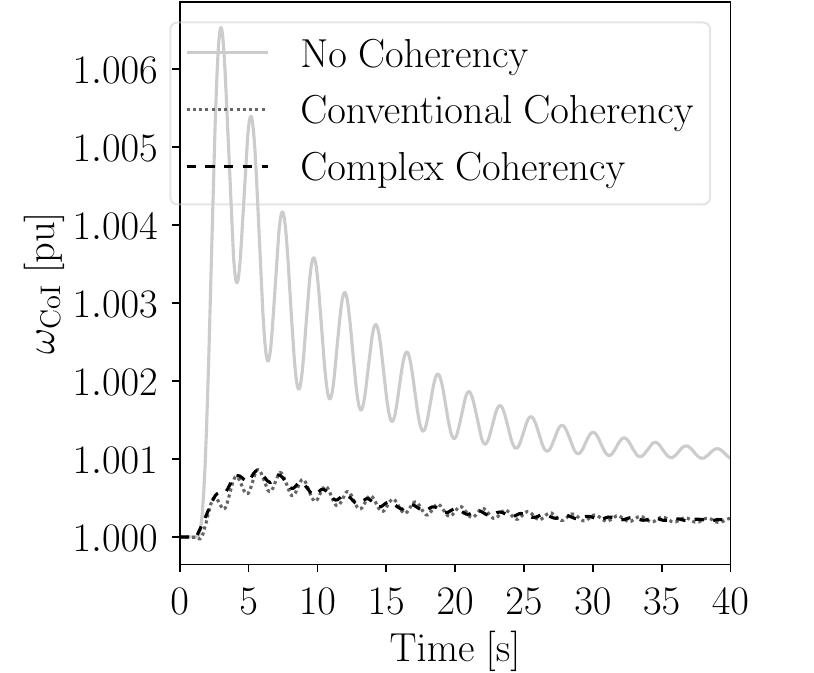}
        \subcaption*{(a)} 
    \end{minipage}\hfill
    \begin{minipage}{0.5\linewidth}
        \centering
        \includegraphics[width=\linewidth]{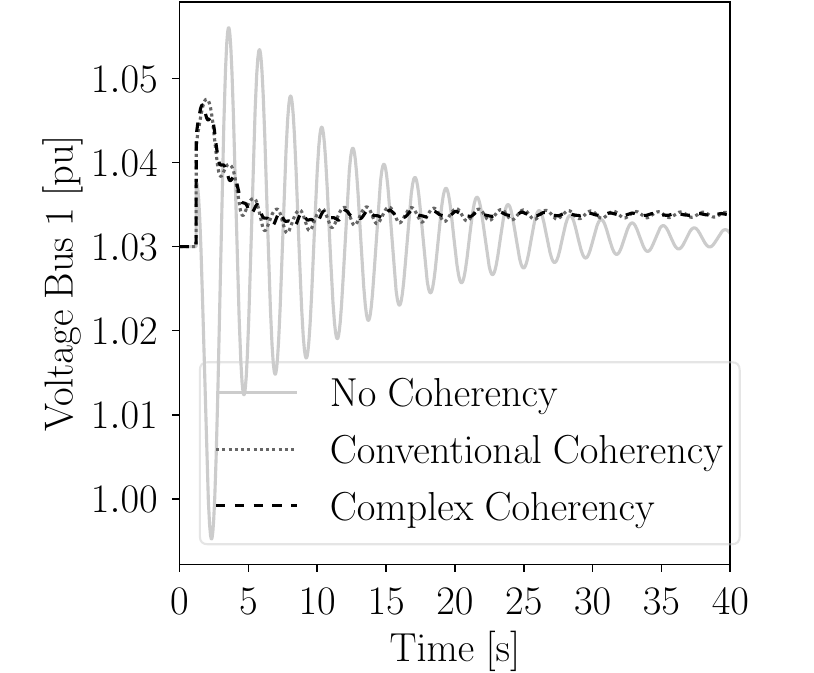}
        \subcaption*{(b)} 
    \end{minipage}
    \caption{Two-area system - Loss of a circuit of the line between buses 7 and 8, $\mathrm{C}=0.25$.  Frequency of the \ac{coi} (a) and voltage at bus 1 (b).}
    \label{fig:conv_vs_cmplx}
\end{figure}

In this section, we compare conventional coherency---defined as a relationship between signal frequencies or phases---and its generalized dynamic definition based on the \ac{cf} concept, as formalized in (\ref{eq:cc}).  In this generalized framework, the coherency function incorporates not only the imaginary component of the signal frequency but also its real part, which relates to variations in the signal magnitude.
Figure~\ref{fig:conv_vs_cmplx} displays the frequency of the \ac{coi} and the voltage at bus 1 following the loss of the line connecting buses 7 and 8.  

Three cases are simulated: No Coherency, with no additional coherency control applied to the generators ($C=0$); Conventional Coherency, which uses only (\ref{eq_curr_prop_imag}) for control, leaving the current magnitude defined by initial conditions; and Complex Coherency, where both magnitude and phase are defined with (\ref{eq_curr_prop_real}) and~(\ref{eq_curr_prop_imag}), respectively.  For the two coherency control cases (Conventional and Complex), the control parameter is set to $C=0.25$ at generation buses 1, 2, and 4.

Both variables---the frequency of the \ac{coi} and the voltage at bus 1---show improved damping when coherency control is applied, whether using the conventional or complex method.  However, the complex coherency achieves better (more damped) performance, as it benefits from a more precise representation of the current dynamics by considering both magnitude and phase.  This results is consistent with the study presented in \cite{rhocomp} that shows that the real part of the \ac{cf} is effective to damp oscillations. 

\subsection{39-Bus system}
The IEEE 39-bus system has historically been used to study coherency among \acp{sm} and identify the coherent areas they form.  The system comprises 10 \acp{sm} with very high inertia constants, as each one may actually represent a dynamic equivalent of an entire sub-area.  The \ac{sm} located at bus 39 is the largest, and due to its high inertia, it is typically segregated into its own unique cluster.  Other standard clusters, identified through electrical distance and dynamic analysis \cite{coherencydef}, include \acp{sm} 2 to 7, 10 and 8, and finally \ac{sm} 9, which is isolated on a radial branch.  These sets arise naturally due to the system's dynamics and topology.

In this section, we aim at forcing coherency between devices that are not part of the same natural cluster.  Specifically, we target machines that are the most electrically distant and inherently belong to different dynamic groups.  To this end, an \ac{ibr} is included at bus 36 with coherency control replacing \ac{sm} 7 and designed to imitate the current dynamics of \ac{sm} 1.  These two machines are among the most electrically distant in the system and therefore represent an interesting case for establishing forced coherency.

All \acp{sm} are modeled using a 4th-order two-axis representation, each equipped with an IEEE Type AC4 AVR and a Type 2 PSS.

\begin{figure}[htbp]
    \centering
    \begin{minipage}{0.5\linewidth}
        \centering
        \includegraphics[width=\linewidth]{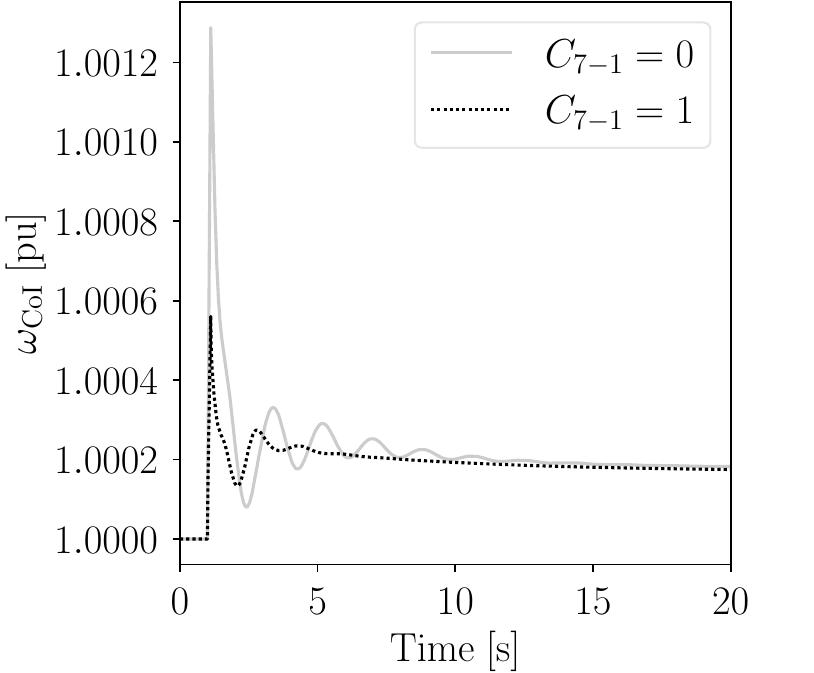}
        \subcaption*{(a)} 
    \end{minipage}\hfill
    \begin{minipage}{0.5\linewidth}
        \centering
        \includegraphics[width=\linewidth]{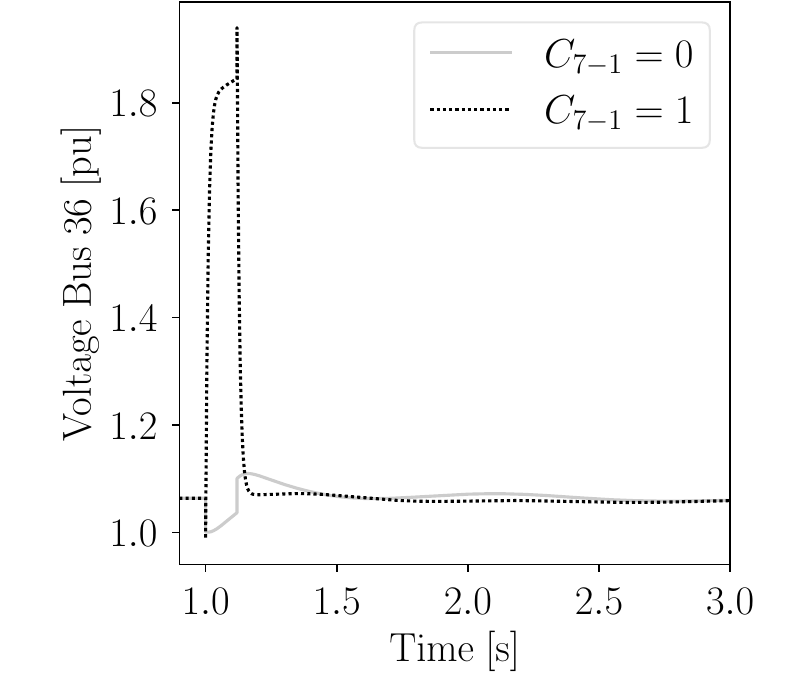}
        \subcaption*{(b)} 
    \end{minipage}
    \caption{IEEE 39-Bus system – Three phase fault at bus 1 - Frequency of the \ac{coi} (a) and voltage at bus 36 (b).}
    \label{fig:IEEE39_1}
\end{figure}

Figure~\ref{fig:IEEE39_1} displays the frequency of the \ac{coi} and the voltage at bus 36 during a three-phase fault at bus 1, cleared after 120 ms.  Two cases are presented: the first, denoted as $C_{7-1}=0$, represents the base case without additional control; the second, denoted as $C_{7-1}=1$, considers the replacement of \ac{sm} 7 with an \ac{ibr} whose coherency control mimics the current dynamics of \ac{sm} 1.

As expected, system damping increases significantly, and the overall dynamic response is improved.  This enhancement is clearly observed in the frequency of the \ac{coi}.  However, the local effect of the three-phase fault near bus 1 is propagated by the coherency control, impacting on the voltage at bus 36.  The control overcompensates, creating a response as if the short-circuit fault were electrically close to that bus.  This may cause unwanted dynamic behavior of the control and potentially leads to current or voltage limitations.

\begin{figure}[htbp]
    \centering
    \begin{minipage}{0.5\linewidth}
        \centering
        \includegraphics[width=\linewidth]{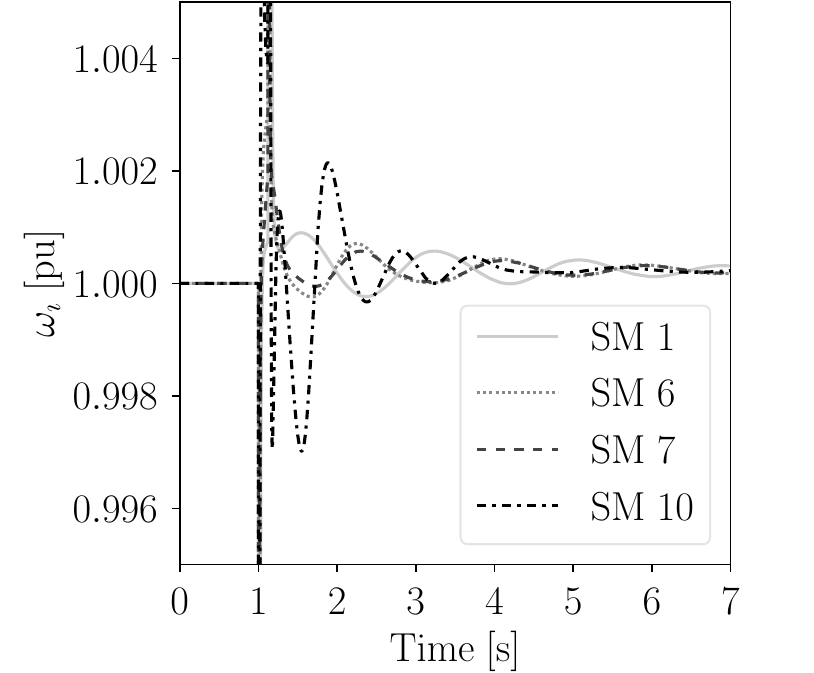}
        \subcaption*{(a) $\omega_{\imath}:\, C_{7-1}=0$.}
    \end{minipage}\hfill
    \begin{minipage}{0.5\linewidth}
        \centering
        \includegraphics[width=\linewidth]{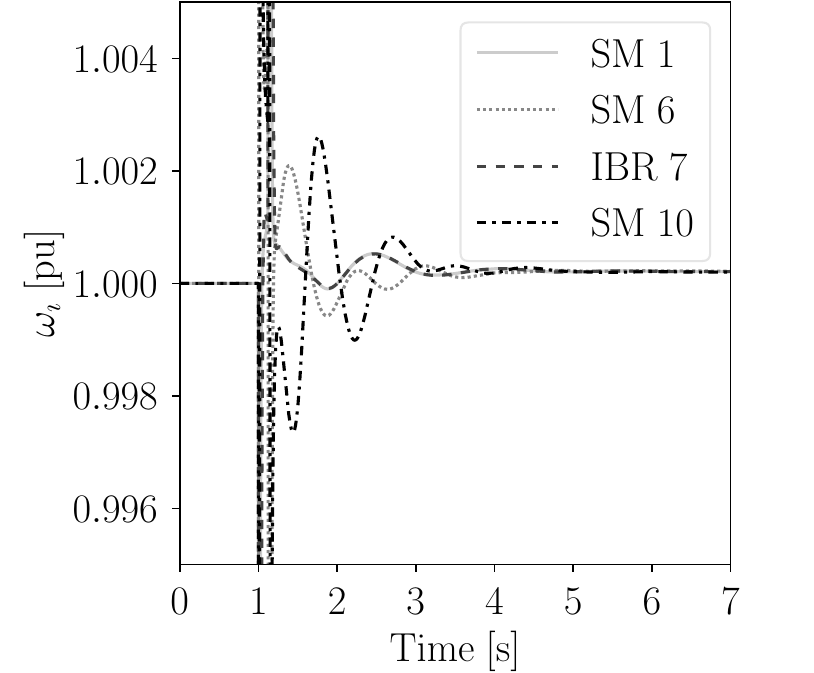}
        \subcaption*{(b) $\omega_{\imath}:\, C_{7-1}=1$,}
    \end{minipage}
    \centering
    \begin{minipage}{0.5\linewidth}
        \centering
        \includegraphics[width=\linewidth]{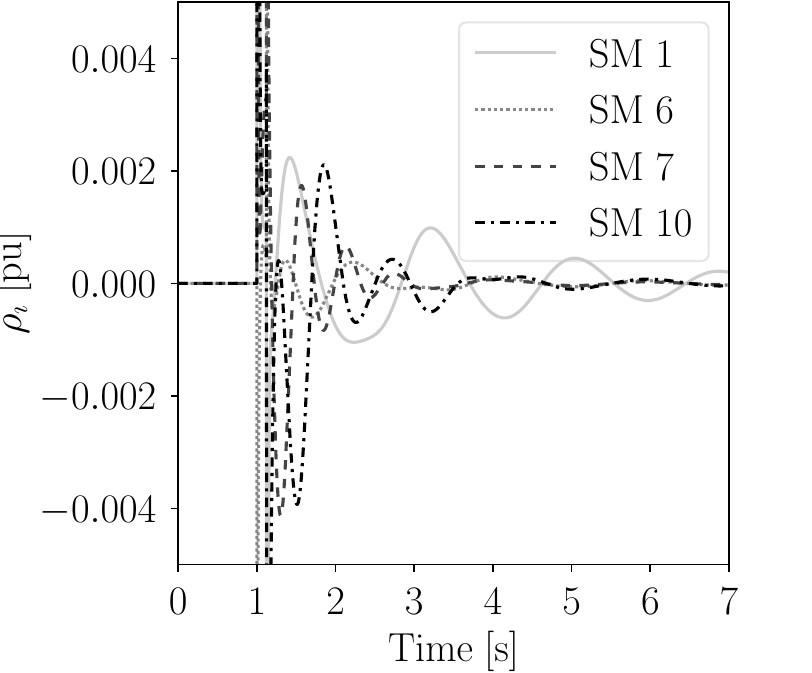}
        \subcaption*{(c) $\rho_{\imath}:\, C_{7-1}=0$.}
    \end{minipage}\hfill
    \begin{minipage}{0.5\linewidth}
        \centering
        \includegraphics[width=\linewidth]{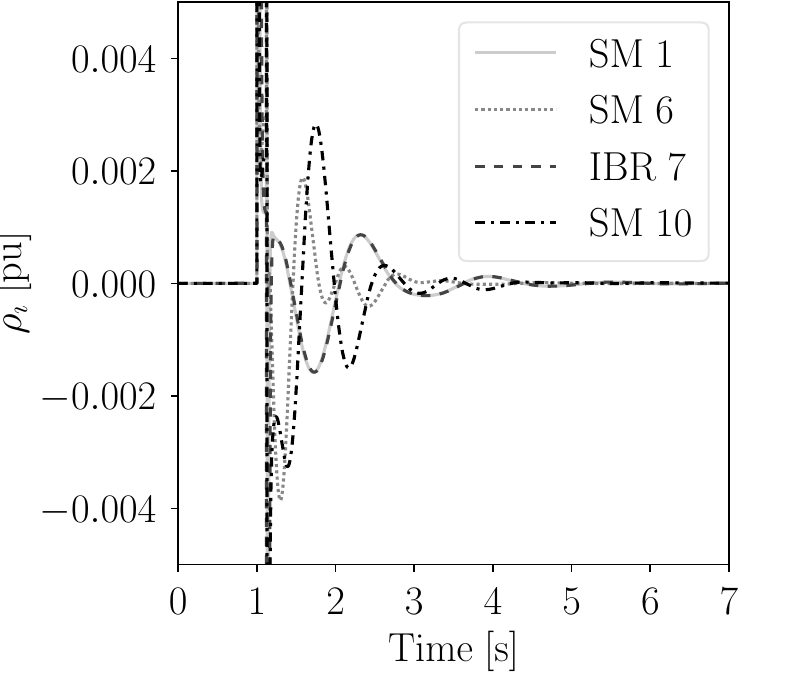}
        \subcaption*{(d) $\rho_{\imath}:\, C_{7-1}=1$,}
    \end{minipage}
    \caption{IEEE 39-Bus system – Three phase fault at bus 1 - real and imaginary parts of the \ac{cf} of the current of generators 1, 6, 7 and 10 for $C_{7-1}=0$  and $C_{7-1}=1$.}
    \label{fig:dyn_clust}
\end{figure}

Figure~\ref{fig:dyn_clust} illustrates the dynamic clustering by showing the real and imaginary parts of the \ac{cf} of the current for generators 1, 6, 7, and 10.  These variables, estimated through a PLL, are shown following a three-phase fault at bus 1 for both the $C_{7-1}=0$ and $C_{7-1}=1$ cases, indicating how the generators are grouped under each condition. Without any additional control (panels a and c), the \ac{cf} of the injected currents appears non-coherent in the first two seconds following the fault.  Subsequently, they begin to oscillate relative to \ac{sm} 1.  In particular, \ac{sm} 6 oscillates almost entirely in anti-phase with \ac{sm} 1.  However, when coherency control is included replacing \ac{sm} 7 to imitate the current dynamics of \ac{sm} 1 (panels b and d), a very different dynamic behavior is observed.  For example, the clear anti-phase oscillation between \acp{sm} 6 and 1 is no longer present.  By establishing coherency between the current dynamics injected at buses 36 and 39 as a control objective, the generators electrically close to bus 36 (e.g., \ac{sm} 6) are forced to behave more coherently with \ac{sm} 1.  This also reduces the time required for the system to reach a steady state.

\section{Conclusion and Future Work}\label{sec:conclusion}
This paper proposes a coherency control strategy for \acp{ibr} that directly regulates the degree of coherence with a remote reference source.  The proposed control is derived by formulating coherency through the equality of the complex frequency (\ac{cf}) of injected currents, enforcing a proportional magnitude and a constant phase shift relative to a measured reference current, making the controller technology-agnostic.

Case studies based on the two-area and IEEE 39-bus systems demonstrate the controller's potential.  The coherency parameter $C$ enables a smooth transition to coherent operation, effectively dampening power oscillations and reducing the settling time as it increases.  Furthermore, incorporating the real part of the \ac{cf} provides additional dynamic improvements over conventional frequency-based coherency.  Nevertheless, the analysis also reveals that coherency control can propagate local disturbances, and its performance is subject to practical limitations such as communication delays and measurement noise, which impose constraints on the feasible distance for implementation.

Future work will focus on enhancing the robustness and practicality of the control.  Key directions include: integrating current and voltage limitation algorithms; exploring the implementation within \ac{gfm} inverter architectures; incorporating support for \ac{wac} to improve system-wide coordination; and developing weighted reference schemes that depend on multiple generators or virtual references to mitigate the risk of propagating local faults.

\end{document}